\documentclass{winnower}
\usepackage{lineno,hyperref}
\modulolinenumbers[5]
\usepackage{float}
\usepackage{amsmath}
\begin{document}

\title{\textbf{Distribution specificities of long-period comets\textquoteright perihelia. Hypothesis of the large planetary body on the periphery of the Solar System}}

\author{Ayyub Guliyev}
\author{Rustam Guliyev}
\affil{Shamakhy Astrophysical Observatory, Azerbaijan National Academy of Sciences, ShAO Y. Mammadaliyev, AZ-1000, Baku, Azerbaijan; rustamdb@gmail.com, quliyevayyub@gmail.com}

\date{}

\maketitle

\begin{abstract}
The present paper reviews selected aspects of the Guliyev\textquoteright s hypothesis about the massive celestial body at a distance of 250--400 AU from the Sun as well as the factor of comets transfer. The analysis covers 1249 comets observed up to 2017, having perihelion and aphelion distances greater than 0.1 and 30 AU respectively. It is shown, that the conjecture of the point around which cometary perihelia might be concentrated, is not consistent. On the issue of perihelia distribution, priority should be given to the assumption that there is a plane or planes around which the concentration takes place. The search engine for such planes was applied to numerous cometary groups, separated by clusters T (discover date), $e$, $q$, $H$ (absolute magnitude), $Q$, $1/a_{ori}$, etc. A total of 24 comet groups were investigated. In almost all cases there are detected two types of planes or zones: the first one is very close to the ecliptic, another one is about perpendicular to it and has the parameters: $i_p = 86^{\circ}$, $\Omega_p = 271.7^{\circ}$. The existence of the first area appears to be related to the influence of giant planets. The Guliyev\textquoteright s hypothesis says that there is a massive perturber in the second zone, at a distance of 250--400 AU. It shows that number of aphelia and distant nodes of cometary orbits in this interval (within statistical confidence) significantly exceeds the expected background. Analysis of the angular parameters of the comets, calculated relative to the second plane (reference point is the ascending node of a large circle) displays clear patterns: shortage of comets near $i' = 180^{\circ}$, excess of them near $B’= 0^{\circ}$ (ecliptic latitude of perihelion) and shortage near $B’=-90^{\circ}$. The analysis also shows irregularity of distant nodes, overpopulation of perihelion longitudes in the range $350^{\circ}$--$20^{\circ}$. Plotted distributions of aphelia N($Q$) and distant cometary nodes clearly indicate a perturbation of the natural course near 300 AU. On the basis of collected cometary data, we have estimated orbital elements of the hypothetical planetary body:
$$ a = 337\, \mathrm{AU};\: e = 0.14;\: \omega = 57^{\circ};\: \Omega = 272.7^{\circ};\: i = 86^{\circ}$$
Naturally, each value may contain some errors.

In order to test the stability of such an orbit, the planet was integrated  for $10^7$ yr, assuming that its mass is about $\sim10M_{\oplus}$. The orbits of 33 comets (having aphelia and distant nodes 286--388 AU) are also integrated in the past for a million years in order to trace possible dynamic relationship with the planet. In doing so, we varied the mean anomaly of the planet from $0^{\circ}$ to $360^{\circ}$ by $10^{\circ}$ in each cycle of numerical explorations. A number of close encounters between comets and the hypothetical body have been discovered.
\end{abstract}

\section{Introduction}

The study of comets is essential for understanding the formation and evolution of the Solar System. Comets contain primordial substance as well as being indicators of space. A study of cometary perihelia distribution has a long history. Historical reviews on the subject were provided by \cite{lyttleton1953comets,vsekhsvyatskij1966origin,radzievskii1987origin} et al. For a long time, \cite{radzievskij1980some,radzievskii1977cometary} had focused on that issue. The issue is important because one of the keys to finding solution to the cometary problem might be hiding out there. The complexity of the issue is that the distribution is affected by the condition of comets appearance or the observational selection effects affecting their discovery \cite{holetschek1891ueber}. The latter is particularly evident in the fact that the number of comets discovered in the north and south is not the same, comets with large perihelion distances are being discovered principally near the ecliptic \cite{yabushita1979statistical}, etc. Probably this should also include complexities associated with the discovery of comets in the galactic plane.

The concentration of cometary perihelion positions near the ecliptic is unique to periodic comets. There is no straightforward concentration of LPCs perihelia were observed in that area. Some analysts considered that the concentration of cometary perihelia takes place near the galactic plane \cite{oja1975perihelion,radzievsky1979regularities}. There is an attempt of Radzievskii et al., to prove that the perihelia of LPCs are concentrated towards the solar apex (peculiar solar motion). This attempt was challenged in \cite{guliev1985version}. There is also an article of \cite{natanson1923origin} who focused on the search for a possible point of perihelia concentration on the Celestial Sphere. There are investigations (\cite{guliev1992possibility,guliyev2007relationship}) related to the influence of the trans-Neptune planetary bodies on the perihelia distribution of the individual cometary groups. These issues, based on the most recent data and some new approaches will be the survey items of the present paper.

\section{Used data}

This research covers the comets with the following characteristics:

\begin{itemize}
	\item Comets, discovered up to 2017
	\item Comets having perihelion distances greater 0.1 AU (therefore, short-perihelion comets are not considered)
	\item Comets with aphelion distances greater 30 AU, i.e. larger than Neptune's orbital distance 
	\item If the comet has been splitted, only the data of a larger fragment of its nuclei labeled A are used
\end{itemize}

Compiled in this way, our basic table contains data of 1249 comets. It begins with the comet 1P/-239 K1 and ends with the object C/2016 X1. While data compilation, we have used the catalogue \cite{marsden2008catalogue} and selected issues of the Minor Planet Electronic Circulars (MPEC) for the period 2008--2016. The restriction on the parameter $Q$ is introduced due to the possible relationship between some comets and KBOs (\cite{guliev1987possible,guliev2007transneptunian}). Otherwise we could place constraints on periods $P > 200$ yr, how it's done in numerous works on cometary statistics. But in that case, we would have to use data of only 1199 comets.

During the workflow, this sample will be separated into individual groups  from time to time to be analyzed.

\section{Computing algorithms}

Let's say we are looking for a plane (great circle on the celestial sphere), near which the concentration of cometary perihelia takes place. The position of such a plane is determined by the method described in \cite{guliev1989transplutonian}. First, we need to determine the directional cosines of the radius vector of each perihelion point in a Cartesian ecliptic coordinate system. Further, by the method of least squares, we derive empirical expressions for the three planes around which the points have a minimum variance. They are presented as follows:
$$Z = ax+by;\: Y=az+bx;\: X=bz+ay$$

To facilitate the task, the second and third expressions are also expressed in terms of $Z$, since it signifies a deviation from the ecliptic. In the next stage, preference is given to an empirical expression issuing the minimum sum of $Z^2$, which we designate as $S$ with subscript notation. According to the case with minimum variance we set the Types of plane: 1, 2 or 3. The plane is defined by two parameters: $\Omega_p$---longitude of the ascending node of the corresponding great circle on the celestial sphere and $i_p$---the inclination of this circle to the ecliptic. In advance, we note that we have not encountered cases in the calculations, when the plane Type 2 seemed to be dominant within the framework. In practice, it is always close to the plane Type 1, but with a greater $S$. Therefore, it will be mentioned rarely.

At this stage of the calculations, we applied the LINEST function from the MS Excel software environment. It returns not only the required coefficients, but also their errors, determinacy level of the obtained empirical expression and the residual variance $S$. The latter is particularly valuable for comparative analysis in the case of different planes.

There may be used more advanced mathematical methods for investigating distributions of cometary perihelia. At least, these techniques had proved to be quite effective where they were applied.

\section{Results of calculations}

Application of the above described algorithm to the basic table shows the existence of the plane Type 1, near which there is a cluster of perihelia

\begin{equation} \label{pln1}
i_p = 86.18^{\circ};\: \Omega_p = 271.74^{\circ};\: (S = 368.18)
\end{equation}

Moreover, another one plane was discovered (Type 3)
\begin{equation} \label{pln2}
i_p = 4.67^{\circ};\: \Omega_p = 303.71^{\circ};\: (S = 374.85)
\end{equation}

It is not difficult to identify the nature of the last one, since it is close to the ecliptic. However, the meaning of the first one is far from clear. In particular, it does not match the galactic plane:$\:i_p = 60.2^{\circ};\: \Omega_p = 269.3^{\circ}$. By comparison, note there are 146, 149 and 88 of perihelia in the $5^{\circ}$ vicinity of the planes \eqref{pln1}, \eqref{pln2} and the galactic plane, respectively. The dominance of the galactic plane over the \eqref{pln1} and \eqref{pln2}, takes place if the $5^{\circ}$ limit is significantly enlarged. For example, on the basis of $45^{\circ}$ vicinity, these numbers look like 943, 924 and 994, respectively. In that case, however, misleading to speak of statistical confidence. 

The technique used, makes it possible to identify errors in the parameters $\Omega_p$ and $i_p$. Taking this into account, it can be argued that parameters of the dominant plane Type 1 correspond to the intervals:
$$270.3^{\circ} < \Omega_p < 272.8^{\circ};\: 84.4^{\circ} < i_p <87.7^{\circ}$$

Similarly, for \eqref{pln2}, the following intervals are found:
$$297.8^{\circ} < \Omega_p < 305.9^{\circ};\: 2.55^{\circ} < i_p <6.79^{\circ}$$

In order to look into the reasons for the existence of the plane \eqref{pln1}, we decided to analyze the distribution of perihelia individually regarding various comet groups:
\begin{itemize}
	\item Comets with eccentricities distinct from 1 (N = 825)
	\item Comets with known \textquotedblleft original\textquotedblright orbits (so-called initial orbital elements) (N = 499) \cite{marsden2008catalogue}
	\item Comets with perihelion distances greater than 7 AU (N = 30); 6 AU (N = 63); 5 AU (N = 117); 4 AU (N = 180); 3 AU (N = 288); 2 AU (N = 462); 1 AU (N = 798) and less than 1 AU (N = 441)
	\item Comets discovered prior to 1750 (N = 101); 1800 (N = 138); 1850 (N = 197); 1900 (N = 327); 1950 (N = 452), 2000 (N = 738)
	\item Comets with aphelion distances less than 6 AU (N = 297); from 30 to 250 (N = 170), from 30 to 200 AU, from 30 to 150 AU, from 30 to 100 AU, from 30 to 50 AU, more than 1000 AU, excluding hyperbolics (N = 638)---these comets are not included in our sample, they are used only for comparison
	\item 3 groups of comets with absolute magnitude less than $9^m$, from $6^m$ to $9^m$, and brighter than $6^m$ (N = 99, 198 and 220, respectively) \cite{vsekhsvyatskij1958physical}
	\item Comets with osculating hyperbolic eccentricities (N = 296)
	\item Comets with some other characteristic features
\end{itemize}

Table ~\ref{tab:1} shows the search results for optimal planes for the cometary groups, separated by perihelion distances. The data of this table show that existence of the plane Type 1 is particularly inherent to the comets with $q < 1$ AU. In all other cases, two values of the $S$ parameter are close to each other. Still in these cases, the computed planes of Type 1 are very similar to \eqref{pln1}.  The observational selection effects (comets with large perihelia are discovered basically near the ecliptic) which artificially increases the dominance of planes Type 2, should also be taken into account.

It was also estimated that the dominance of plane Type 3 over Type 1 is inherent in comets with $Q < 10000$ AU (484 comets).

\begin{table}[!h] 
	\centering
	\caption{Inclination, longitude of the ascending node of planes, near which the cometary perihelia are concentrated}
	\label{tab:1}
	\begin{tabular}{lllll}
		\hline
		Group feature      & $i_p$(deg)& $\Omega_p$(deg)& N& $S_1$ \& $S_3$ \\ \hline
		All 1256 comets    & 86.1 & 271.8 & 1249 & 370.17   \\
		& 4.7  & 303.3 &      & 375.8    \\
		$q>7$ AU 		   & 82.6 & 266.8 & 30   & 10.98    \\
		& 15.1 & 200.6 &      & 8        \\
		$q>6$ AU           & 84.2 & 272.2 & 62   & 20.3     \\
		& 5.6  & 258.2 &      & 19.2     \\
		$q>5$ AU           & 84.5 & 271.5 & 117  & 34.15    \\
		& 5.9  & 245.3 &      & 33.3     \\
		$q>4$ AU           & 86.7 & 271.7 & 178  & 54.76    \\
		& 4.6  & 225.5 &      & 54.37    \\
		$q>3$ AU           & 88.8 & 269.8 & 287  & 87.52    \\
		& 2.4  & 211.0 &      & 87.03    \\
		$q>2$ AU           & 89.2 & 270.3 & 461  & 137.87   \\
		& 1.0  & 309.3 &      & 137.65   \\
		$q>1$ AU           & 85.9 & 271.6 & 803  & 239.75   \\
		& 4.2  & 290.6 &      & 226.94   \\
		$q<1$ AU           & 85.8 & 272.0 & 446  & 127.41   \\
		& 6.3  & 311.6 &      & 144.81   \\ \hline
	\end{tabular}
\end{table}

Table \ref{tab:2} shows the results of calculations for comets discovered prior to a certain epoch (up to 1750, 1800, 1850, 1900, 1950, 2000).

\begin{table}[H]
	\centering
	\caption{Inclination, longitude of the ascending node of the planes with minimal $S$ for comets, separated by epoch of discovery.}
	\label{tab:2}
	\begin{tabular}{lllll}
		\hline
		Up to & $i_p$(deg)   & $\Omega_p$(deg) & N   & $S_1$ \& $S_3$ \\ \hline
		1750  & 87.2 & 276.4 & 101 & 29.93    \\
		& 8.4  & 341   &     & 29.77    \\
		1800  & 86.7 & 274.2 & 138 & 40.41    \\
		& 6.6  & 327.6 &     & 42.97    \\
		1850  & 85.4 & 271.4 & 197 & 58.54    \\
		& 6.1  & 309.1 &     & 60.34    \\
		1900  & 86   & 270.7 & 327 & 94.05    \\
		& 4.4  & 280   &     & 101.73   \\
		1950  & 86   & 272.3 & 457 & 134.69   \\
		& 4.2  & 276.7 &     & 139.58   \\
		2000  & 85.3 & 272.5 & 740 & 216.67   \\
		& 5.5  & 297   &     & 226.57  \\ \hline
	\end{tabular}
\end{table}

As shown in the table, in all cases, planes Type 1 took precedence over planes Type 3 in terms of $S$ parameter. Most important, though, the position of the corresponding planes has remained fairly constant at all times.

Table \ref{tab:3} shows the results of calculations for cometary groups, separated by aphelion distances, eccentricity and absolute magnitude according to the scale of Vsekhsvyatskii \cite{vsekhsvyatskij1958physical}. The table also contain similar results for periodic comets having $Q < 6$ AU. As expected, results provide a plane with a minimum variance, closely approximated to the ecliptic. Even in this case, the plane of Type 1 is inclined only $12^{\circ}$ relative to \eqref{pln1}.

\begin{table}[H]
	\centering
	\caption{Positions of optimal planes for cometary perihelia with various characteristics}
	\label{tab:3}
	\begin{tabular}{lllll}
		\hline
		Group feature       & $i_p$(deg)   & $\Omega_p$(deg) & N    & $S_1$ \& $S_3$ \\ \hline
		$Q>250$ AU & 87.7 & 272.3 & 1082 & 310.1   \\
		& 5.2  & 301   &      & 333.9   \\
		Original            & 88.6 & 271.4 & 499  & 140.1   \\
		& 4.6  & 303.4 &      & 154.1   \\
		$Q>1000$ AU   & 84.8 & 271.5 & 638  & 180.94  \\
		& 6.9  & 306.5 &      & 194.91  \\
		$e>1$      & 88.8 & 272.1 & 296  & 88.81   \\
		& 1.6  & 306.8 &      & 96.57   \\
		$Q<6$ AU        & 69   & 270.3 & 297  & 136.02  \\
		& 1.1  & 212.7 &      & 3.61    \\
		$H<9^m$        & 80.8 & 273.4 & 99   & 27.04   \\
		& 11.4 & 292.3 &      & 31.05   \\
		$6^m<H<9^m$          & 88.8 & 276.2 & 198  & 59.15   \\
		& 1.4  & 273.5 &   & 63.66   \\
		$H<6^m$            & 84   & 269.4 &   220   & 57.51   \\
		& 6.5  & 276   &      & 62.32   \\
		$30<Q<250$ AU      & 88   & 268.2 & 170  & 59.54   \\
		& 2.5  & 326.1 &      & 41.42   \\
		$282<Q<400$ AU        & 81   & 9.2   & 46   & 12.29   \\
		& 9.2  & 274.9 &      & 12.51  \\ \hline
	\end{tabular}
\end{table}

As shown in Table 3, we have separated the comets by absolute magnitude ($H$) into three groups (517 long-period comets were derived from a range of sources) brighter than $6^m$, from $6^m$ to $9^m$ and fainter than $9^m$. Calculations carried out on them, did not lead to unexpected results. In all three cases, we have obtained planes that differ little from (1). We might also note that we have used incomplete data in this analysis, but only accessible from sources (the book of \cite{vsekhsvyatskij1958physical} and its continuations).

Table \ref{tab:3} also shows the results for comets, whose original orbits are contained in the catalogue \cite{marsden2008catalogue}.  It is also established that the plane Type 3 is more suitable for the comets having $Q\in(30,250)$ AU, due to the minimality of $S$. 

\section{On the possibility of perihelia concentration around some point}

Now suppose that we want to investigate the distribution of perihelia relative to a point having ecliptic coordinates $L_0$ and $B_0$ on the celestial sphere. The angular distance $\lambda$ in this case can be defined by formula:
\begin{equation}
\cos\lambda=\sin B\sin B_0+\cos B\cos B_0\cos(L-L_0)
\end{equation}

There are evidences in the scientific literature \cite{oja1975perihelion,radzievskii1977cometary,radzievskij1980some} that perihelia might accumulate towards the solar apex, at ecliptic coordinates $L_A = 27^{\circ}$; $B_A = 53.5^{\circ}$. In order to test this, we computed the angular distance of all 1249 perihelia between the solar apex using formula (3). Applying this to the basic table, we can quantify perihelia in any surrounding of the solar apex. In \cite{guliev1985version} the point ($L_0; B_0$) was replaced by some other points on the celestial sphere for a comparative analysis of perihelion frequency. However, the dominance of the solar apex over the many of chosen directions was not found.

In current work, we come back to the issue once more, on the grounds of grown availability of data and more thorough analysis. In the calculations, we set 12 values of the parameter $L_0$ ($0^{\circ}, 30^{\circ},\dots,330^{\circ}$) and 6 of $B_0$ ($33.56^{\circ},48.19^{\circ}, 60^{\circ}, 70.53^{\circ}, 80.41^{\circ}, 90^{\circ}$). Thus, we have identified 61 equidistant points on the northern hemisphere for a comparative analysis of their surroundings. Next, we determined the quantity of perihelia in the $\lambda$-surrounding for each of them. The calculations were carried out in cases where $\lambda$ is equal to $5^{\circ}, 10^{\circ}, 15^{\circ}, 20^{\circ}, 25^{\circ}$ and $30^{\circ}$. For the analysis, the following values were found: N---the number of perihelia in the $\lambda$-surrounding of the solar apex, $\bar{n}$---average number of perihelia corresponding to the $\lambda$-surrounding of the 61 points on the celestial sphere, $\sigma$---standard deviation, t---standard ratio, which, according to the one-sided Student's t-distribution should be greater than 1.67 at the significance level of 0.05 \cite{gmurman1968fundamentals}. 

\begin{table}[H]
	\centering
	\caption{Comparative frequencies of perihelia relative to the solar apex and other directions}
	\label{my-label}
	\begin{tabular}{lllll}
		\hline
		$\lambda$(deg)   & N   & $\bar{n}$     & $\sigma$     & t    \\ \hline
		50  & 3   & 2.51  & 1.83  & 0.27 \\
		100 & 12  & 10.56 & 5.63  & 0.26 \\
		150 & 33  & 23.61 & 10.15 & 0.93 \\
		200 & 59  & 42.07 & 14.46 & 1.17 \\
		250 & 80  & 65.79 & 18.4  & 0.77 \\
		300 & 107 & 95.52 & 21.24 & 0.54 \\ \hline
	\end{tabular}
\end{table}

As shown in Table 4, in no case, the expected value of t is achieved. Continue the calculations for the southern hemisphere does not make sense, since in this case t-parameter will be even smaller. As for the method used by \cite{natanson1923origin}, it does not give grounds for determining the significance of the perihelia concentration relative to the point found through this method.

It may therefore be concluded that the concentration of the LPCs perihelia in the area of solar apex is either non-existent or its significance does not satisfy the requirements of mathematical statistics.

\section{Distribution of the angular orbital elements of long-period comets around the plane Type 1}

To interpret the existence of the plane Type 1, the distribution of the angular orbital elements should be analyzed with respect to this plane. To this end, we calculated the values of $\omega',\, \Omega_d',\, i',\, L'$, and $B'$, using the transition formulas of spherical astronomy, provided that the ascending node of a plane or a great circle outlines the reference point. To determine the statistical confidence of the excess or  deficit of values on some interval, we apply Student's t-test \cite{gmurman1968fundamentals} (hereafter t-test). For independent review, each distribution is accompanied by a set of dispersion values: $n,\, \sigma$, t and $\alpha$ (confidence level).

Table \ref{tab:5} shows the distribution of orbital inclinations of the LPCs. When compiling the table, the range of $i'$ was defined on the principle of equidistant orbital poles. The distribution is described by the parameters: $n = 124.9;\: \sigma = 12.24$. t-values are given in the fourth line. In general, we can see that the distribution is mainly characterized by a minimum in the range $143.1^{\circ}$--$180^{\circ}$. In other ranges, this parameter is distributed quite uniformly. The use of t-test shows that the \textquotedblleft deficiency\textquotedblright of values in this interval has a confidence level $\alpha > 0.95$.  

The values of $B$ and $L$ can be obtained from:
\begin{equation}
\begin{aligned}
L &=\Omega+\arctan (\tan\omega\cos i) \\
B &=\Omega+\arcsin(\sin\omega\sin i)
\end{aligned}
\end{equation}

\begin{table}[H]
	\centering
	\caption{Distribution of the orbital inclinations of 1249 LPCs with respect to the plane Type 1 (n -- indicates discrete uniform distribution)}
	\label{tab:5}
	\begin{tabular}{llll}
		\hline
		$i'$(deg)    & N   & n     & t     \\ \hline
		36.9  & 137 & 124.9 & 0.99  \\
		53.1  & 126 & 124.9 & 0.09  \\
		66.4  & 124 & 124.9 & -0.07 \\
		78.5  & 126 & 125   & 0.09  \\
		90    & 131 & 124.9 & 0.5   \\
		102   & 137 & 125   & 0.99  \\
		113.6 & 127 & 124.9 & 0.17  \\
		126.9 & 133 & 124.9 & 0.66  \\
		143.1 & 110 & 124.9 & -1.22 \\
		180   & 98  & 124.9 & -2.2 \\ \hline
	\end{tabular}
\end{table}

The distribution of LPCs $B'$ (ecliptic latitude of perihelion with respect to \eqref{pln1}) is also of considerable interest. The range of this parameter is defined on the principle of equal areas on the celestial sphere. The distribution is described by the parameters: $n=124.9;\: \sigma=24.31$. Values of t are calculated with respect to the uniform distribution and are given in the fourth line. Here, above all, attention is drawn to a small asymmetry relative to the zeroth argument (584 vs. 665). Largely, it is related to the \textquotedblleft deficiency\textquotedblright of perihelia in the region $B'<-53.13^{\circ}$ (t $= -1.72;\: \alpha = 0.95$). The distribution maximum corresponds to the interval $x\in(-11.54, 0)$ and its $\alpha = 0.92$, see Table \ref{tab:6}.

\begin{table}[H]
	\centering
	\caption{Distribution of $B'$ values for 1249 LPCs with respect to the plane Type 1}
	\label{tab:6}
	\begin{tabular}{llll}
		\hline
		$\sin Bؘ'$& $Bؘ'$(deg)  & N   & t     \\ \hline
		-0.8   & -53.13 & 83  & -1.72 \\
		-0.6   & -36.87 & 106 & -0.78 \\
		-0.4   & -23.58 & 117 & -0.33 \\
		-0.2   & -11.54 & 114 & -0.45 \\
		0      & 0      & 163 & 1.56  \\
		0.2    & 11.54  & 141 & 0.66  \\
		0.4    & 23.58  & 160 & 1.44  \\
		0.6    & 36.87  & 116 & -0.37 \\
		0.8    & 53.13  & 125 & 0     \\
		1      & 90     & 124 & -0.04 \\ \hline
	\end{tabular}
\end{table}

Table \ref{tab:7} shows the distribution of 1249 LPCs distant nodes longitude ($\Omega'_d$) with respect to the plane Type 1. It is specified by a significant minimum and maximum in the range $60^{\circ}$--$90^{\circ}$ (58 values) and $180^{\circ}$--$240^{\circ}$ (149 values), respectively. Its confidence level exceeds 0.95. One possible interpretation of these extremes will be given below.

\begin{table}[H]
	\centering
	\caption{Distribution of distant nodes longitude for 1249 LPCs with respect to the plane Type 1}
	\label{tab:7}
	\begin{tabular}{lll}
		\hline
		$\Omega'_d$(deg) & N   & t     \\ \hline
		30   & 130 & 1.05  \\
		60   & 89  & -0.61 \\
		90   & 58  & -1.87 \\
		120  & 77  & -1.1  \\
		150  & 100 & -0.17 \\
		180  & 120 & 0.65  \\
		210  & 149 & 1.83  \\
		240  & 94  & -0.41 \\
		270  & 91  & -0.53 \\
		300  & 109 & 0.2   \\
		330  & 112 & 0.32  \\
		360  & 120 & 0.65 \\ \hline
	\end{tabular}
\end{table}

Statistically significant extremes were not found in the distribution of $L'$ (ecliptic longitude of perihelion in the new reference system), so these data are not included in this paper. 

Table \ref{tab:8} provides data on the parameter $\omega'$. Analysis of the parameter shows that this distribution is specified by the maximum in the range  $0^{\circ}$--$30^{\circ}$ (166 values). In other intervals of adequate length, the distribution is close to random. A more detailed calculation using the $30^{\circ}$ range showed that the $\omega'$ distribution maximum corresponds to the interval $350^{\circ}$--$20^{\circ}$, which includes 178 values of this parameter. This is extremely important and will be taken into account in further developments.

\begin{table}[H]
	\centering
	\caption{Distribution of the parameter $\omega'$ for 1249 LPCs with respect to the plane \eqref{pln1}}
	\label{tab:8}
	\begin{tabular}{lll}
		\hline
		$\omega'$(deg)  & N   & t     \\ \hline
		30  & 166 & 2.14  \\
		60  & 137 & 1.14  \\
		90  & 92  & -0.42 \\
		120 & 89  & -0.52 \\
		150 & 81  & -0.8  \\
		180 & 91  & -0.45 \\
		210 & 133 & 1     \\
		240 & 102 & -0.07 \\
		270 & 84  & -0.69 \\
		300 & 77  & -0.93 \\
		330 & 74  & -1.04 \\
		360 & 123 & 0.65 \\ \hline
	\end{tabular}
\end{table}

\section{The hypothesis of the existence of a large TNO on the periphery of the Solar System}

First attempt to interpret the existence of the plane Type 1 was made in \cite{guliev1999results}. It was found that number of LPCs distant nodes have background increase in the range 250--400 AU. Based on this fact, it has been assumed that there may be a large TNO-comet source in the zone ($250 < R < 400$ AU, plane of Type 1). Such a celestial body can \textquotedblleft generate\textquotedblright long-period comets as well as comets with the smaller $Q$. This idea was developed in \cite{guliyev2007relationship} and \cite{guliyev&guliyev2012hypotesis}

After the paper of \cite{guliev1999results} was published, the number of LPCs consistent with the principles of our basic list, have increased by almost 76\%, but the effect was preserved. Moreover, the analysis shows that number of comets with the corresponding $Q$ in the range 250--400 AU have increased noticeably also. According to our basic table, their number is 55, whereas in 2000 there were only 31 of them.

\begin{center}
	\begin{figure}[H]
		\includegraphics[width=1\linewidth]{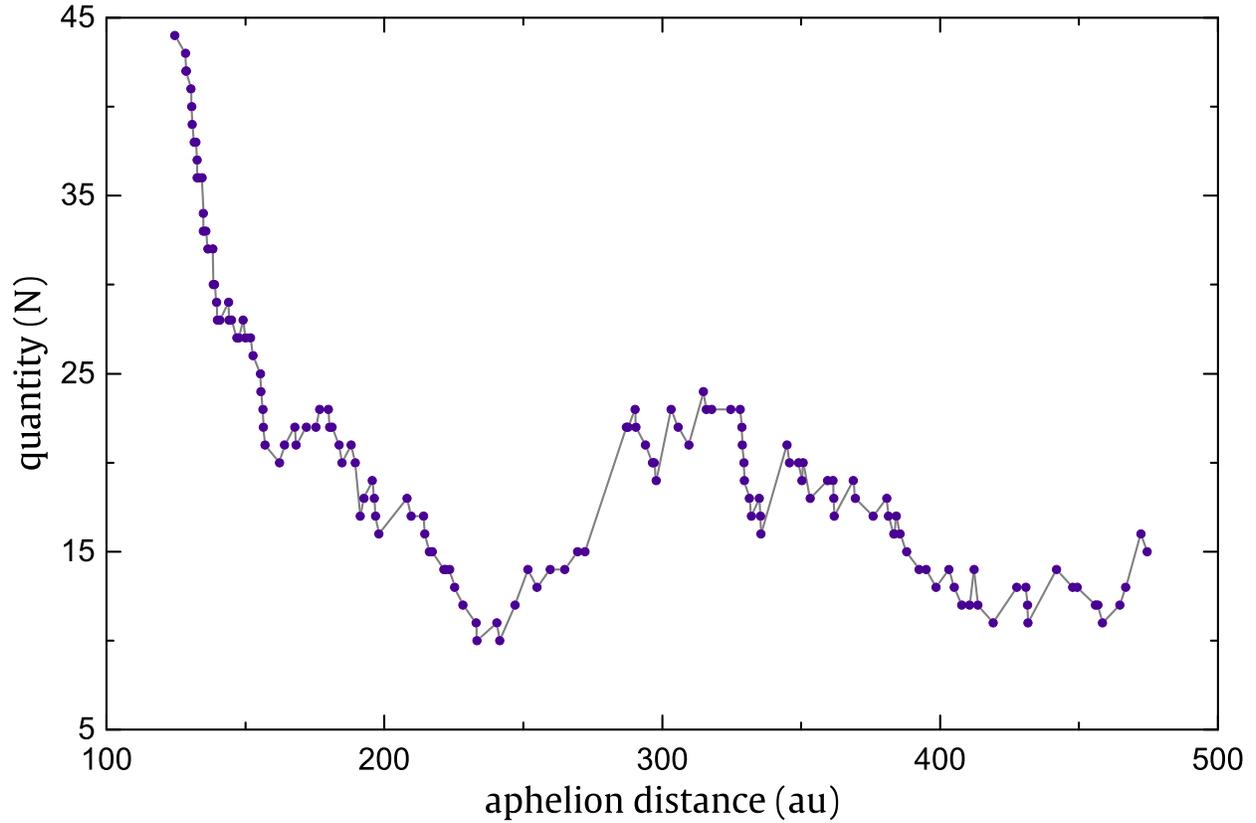}
		\caption{The number of $Q$ at the intertwined intervals for the aphelion distances}
		\label{fig:1}
	\end{figure}
\end{center}

We took 150 comets having $Q\in(100,475)$ AU from our list  and performed one-dimensional analysis on the scale of distances. Using the Sturgess method, we have determined the optimal interval length, $l=47.8$ AU. Adding this value to $Q_i$ each time, we can easily define 150 intervals and calculate the values of the aphelia (N) in each of them. In other words, we applied technique of intertwined intervals in the statistics of cometary aphelia. Figure \ref{fig:1} clearly indicates a deviation from the expected distribution of type $1/Q$ in the range 250--400 AU. This can be another indirect evidence that this area somehow relates to the process of replenishment of the observed LPCs. 

\begin{center}
	\begin{figure}[H]
		\includegraphics[width=1\linewidth]{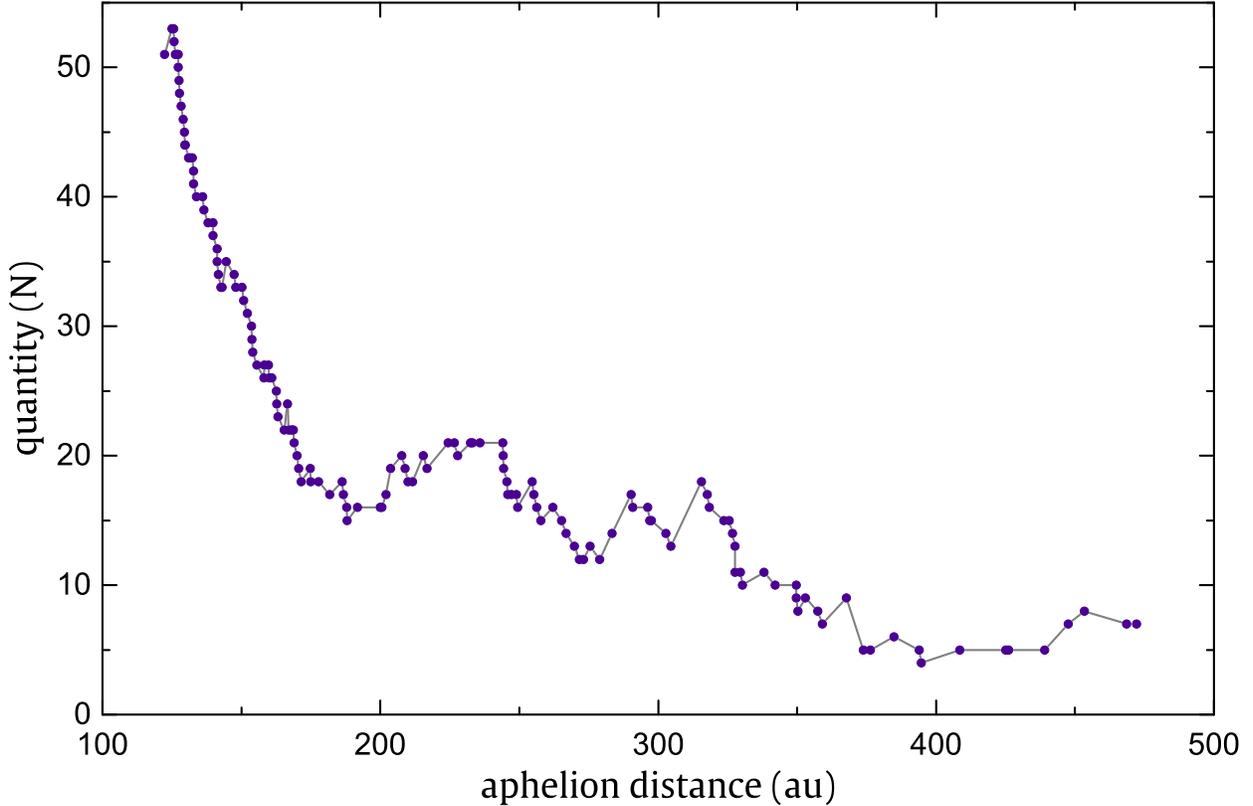}
		\caption{Distribution of aphelion distances of long-period comets in the range 100--500 AU}
		\label{fig:2}
	\end{figure}
\end{center}

A similar but nonetheless important pattern is observed in the distribution of distant nodes relative to the plane \eqref{pln1}. It is shown in Figure \ref{fig:2}. Data on intertwined intervals of 40 AU are also used here. Against that overall background, the maximum deviation is in the range 278.2--318.2 AU, where the number of nodes is equal to 19.

The fact of the \textquotedblleft shortage\textquotedblright of comets with sharply reversed $i'$ can be easily explained within the framework of the hypothesis. To \textquotedblleft generate\textquotedblright such comets, very stringent conditions are required. For the same reason, there is no reverse motion in the system of Jupiter family comets. According to the regularity associated with the distribution of the parameter $\omega'$, the group of comets of an unknown planet has similarities with the Jupiter family. The existence of an overpopulation of the $\sin B'$ around 0 (see Table \ref{tab:6}), as well as its deficit in the range $x\in(-1,-0.8)$, also fits in well with the hypothesis. The distribution of $\Omega'_d$, within the framework of Guliyev's hypothesis, requires additional analysis.

But the main argument of the Guliyev's hypothesis is related to the distribution of distances of LPCs distant nodes. \cite{guliyev2007relationship} demonstrated their overpopulation in the range 250--400 AU, i.e. redundancy relative to some background. Recall that to determine such a background, authors have used data of nodes with respect to 67 regions. In current work, the number of such zones is increased to 229 (the parameter $\Omega_p$ ranged from $0^{\circ}$ to $345^{\circ}$ by $15^{\circ}$, $\sin i_p$ ranged from 0 to 1 by 0.1), and the number of comets is almost 1.5 times (1249 vs. 859) higher than in \cite{guliyev2007relationship}. On the basis of calculations, we have obtained the following dispersion values with respect to the plane \eqref{pln1}:
$$N = 43,\: \bar{n} = 27.51,\: \sigma = 6.34,\: \mathrm{t} = 2.44,\: \alpha > 0.99$$
their description is given in \cite{guliyev2007relationship}.

Varying the applied interval, one can find a more effective value of the t-parameter. For example, in the range 283--357 AU, t $=3.12$, which in fact eliminates the randomness. We note that a similar test was also carried out on the real TNOs, including Eris \cite{guliev2007transneptunian}.

The Guliyev's hypothesis predicts an important effect, which can be verified on the basis of our comet list. It assumes that the hypothetical planet has not only high inclination, but also has a tangible eccentricity of the orbit. Therefore, if we draw an ellipse with the minimum variance along the distant nodes of the corresponding comets and then another one along the aphelion positions in the range 250--400 AU (also with minimum variance), the directions of semi-major axes of these two ellipses should either coincide or deviate minimally from each other. This assumption was verified in \cite{guliyev&guliyev2012hypotesis}, with less data used, and fully aligned with the conjecture.

To test the predictions of the above hypothesis, we selected nine comets having the latitude of lines of apsides within $\pm 5^{\circ}$ with respect to the plane. Table \ref{tab:9} contains the following data of these comets: orbital elements relative to the ecliptic; $Q,\, L',\, B'$---aphelion distance, ecliptic longitude and latitude of the perihelion relative to the plane \eqref{pln1}, where the reference point is the ascending node of the hypothetical planet.

\begin{table}[H]
	\centering
	\caption{Parameters of 9 comets having latitude of perihelion within $\pm 5^{\circ}$ relative to \eqref{pln1}}
	\label{tab:9}
	\begin{tabular}{llllllllll}
		\hline
		Comet     & $q$(AU)     &$e$     & $\omega$(deg)     & $\Omega$(deg)     & $i$(deg)     & $Q$(AU)     & $L'$(deg)    & $B'$(deg)   & $(r-Q)^2$ \\ \hline
		C/2003 L2 & 2.865 & 0.981 & 119.9 & 273.6 & 82.1  & 305.7 & 120.1 & -2.7 & 0.6    \\
		C/2007 M3 & 3.469 & 0.98  & 125.7 & 41.6  & 161.8 & 345   & 14.9  & -1.7 & 3354.6 \\
		C/1920 X1 & 1.148 & 0.994 & 340.9 & 108.8 & 22    & 386.7 & 187   & -1.2 & 1052.1 \\
		C/1911 O1 & 0.489 & 0.997 & 153   & 294.2 & 33.8  & 326.9 & 165.4 & 0.5  & 140.2  \\
		C/1807 R1 & 0.646 & 0.995 & 4.1   & 269.5 & 63.2  & 285.9 & 3.6   & 0.7  & 33     \\
		C/1881 K1 & 0.735 & 0.996 & 354.2 & 272.6 & 63.4  & 362   & 354.7 & 1.3  & 4351.7 \\
		C/2014 S2 & 2.101 & 0.988 & 87.8  & 8.1   & 64.7  & 337.8 & 115.5 & 4    & 1285.9 \\
		C/1846 B1 & 1.481 & 0.992 & 338   & 113.3 & 47.4  & 388.2 & 196.5 & 4.9  & 806.9  \\
		C/2004 F2 & 1.431 & 0.991 & 317.2 & 248.3 & 105   & 305.5 & 318.2 & 5    & 219.9 \\ \hline
	\end{tabular}
\end{table}

As shown in the table, the $Q$ values of the comets are in the range 286--388 AU. If we assume that the planet is moving in this interval, its four estimating elements (semi-major axis, eccentricity and two angular) are: 
$$a = 337\,\mathrm{AU};\:	e = 0.136;\: \Omega = 271.74^{\circ};\: i = 86.18^{\circ}$$

Next, neglecting the values of $B'$ and specifying different values of $\omega$, one can find the optimal ellipse, on which the aphelion positions have the minimum variance. That value was found to be $45^{\circ}$.

A similar task was handled for comets, the distant nodes of which are in the range 286--388 AU relative to the plane \eqref{pln1}. The number of those is 25. Their data are given in Table \ref{tab:10}.

\begin{table}[H]
	\centering
	\caption{Parameters of 25 comets having distant nodes in the range 285--388 AU with respect to \eqref{pln1}}
	\label{tab:10}
	\begin{tabular}{llllllllll}
		\hline
		Comet       & $q$(AU)     & $e$     & $\omega$(deg)     & $\Omega$(deg)     & $i$(deg)     & $R_d$(AU)    & $\Omega'_d$(deg)   & $i'$(deg)    & $(r-R_d)^2$ \\ \hline
		C/1822 N1   & 1.145 & 0.996 & 181.1 & 95.2  & 127.3 & 293.8 & 175   & 146.3 & 662      \\
		C/2004 G1   & 1.202 & 1     & 110.5 & 228.4 & 114.5 & 295.8 & 306.2 & 50.8  & 136      \\
		C/1999 N4   & 5.505 & 1.004 & 90.4  & 345.9 & 156.9 & 296.6 & 202.2 & 87.4  & 42       \\
		C/1748 K1   & 0.625 & 1     & 245.7 & 36.6  & 67.1  & 301.8 & 119.3 & 120   & 3740     \\
		C/2006 CK10 & 1.752 & 0.992 & 143.5 & 243.8 & 144.3 & 303.8 & 342   & 62.6  & 815      \\
		C/2010 A4   & 2.738 & 0.99  & 271.7 & 346.7 & 96.7  & 304.9 & 82    & 75.5  & 5863     \\
		C/2010 J3   & 2.249 & 1     & 180.4 & 101.1 & 14.6  & 305.9 & 182.4 & 100.6 & 74       \\
		C/2014 M1   & 5.545 & 1     & 337.6 & 234.8 & 160.2 & 305.9 & 168   & 78    & 351      \\
		C/1847 N1   & 1.766 & 0.999 & 91.5  & 340.4 & 96.6  & 307.9 & 261.5 & 69.3  & 256      \\
		C/2016 Q2   & 7.199 & 0.982 & 84    & 322.3 & 109.6 & 308.6 & 242.8 & 54.9  & 295      \\
		C/2005 N1   & 1.125 & 0.998 & 80    & 3.2   & 51.2  & 316.2 & 308.8 & 88.8  & 52       \\
		C/1911 O1   & 0.489 & 0.997 & 153   & 294.2 & 33.8  & 320.3 & 345   & 55.4  & 208      \\
		C/1914 M1   & 3.747 & 1.003 & 14    & 271.5 & 71    & 327.9 & 180.8 & 15.1  & 152      \\
		C/1890 O1   & 0.764 & 1     & 85.7  & 15.8  & 63.4  & 327.9 & 298.1 & 100.8 & 607      \\
		C/2006 S3   & 5.131 & 1.003 & 140.1 & 38.4  & 166   & 328.5 & 191.4 & 102   & 384      \\
		C/2002 J5   & 5.727 & 1.001 & 74.8  & 314.1 & 117.2 & 331.3 & 230.1 & 51.3  & 1445     \\
		C/1980 E1   & 3.364 & 1.057 & 135.1 & 114.6 & 1.7   & 335.6 & 180.6 & 87.7  & 400      \\
		C/2007 M3   & 3.469 & 0.98  & 125.7 & 41.6  & 161.8 & 337.3 & 194.4 & 105.3 & 911      \\
		C/1997 A1   & 3.157 & 1.002 & 40    & 135.8 & 145.1 & 345.9 & 333.3 & 117.7 & 414      \\
		C/2008 E3   & 5.531 & 0.998 & 218.1 & 105.7 & 105.1 & 352.1 & 130.3 & 162.3 & 7        \\
		C/2007 D3   & 5.209 & 0.992 & 309.1 & 148.4 & 45.9  & 354.5 & 39.8  & 110.3 & 412      \\
		C/1907 T1   & 0.983 & 1     & 294.4 & 55.9  & 119.6 & 363.1 & 48.7  & 137.4 & 246      \\
		C/1920 X1   & 1.148 & 0.994 & 340.9 & 108.8 & 22    & 372.2 & 6.6   & 107.2 & 397      \\
		C/1914 S1   & 0.713 & 0.999 & 270.3 & 1.6   & 77.8  & 372.9 & 102.1 & 89    & 1        \\
		C/2011 C1   & 0.883 & 0.997 & 84.5  & 192.4 & 16.8  & 386.8 & 196.6 & 83.3  & 6539 \\ \hline    
	\end{tabular}
\end{table}

In addition to the elements referred to the ecliptic, here are: $R_d$---heliocentric distance of the distant node, $\Omega'_d$---its angular value, $i'$---inclination of the cometary orbit in the new frame and $(r - R_d)^2$---squared residual with respect to the obtained ellipse. The new frame means the plane \eqref{pln1}, provided that the reference point is the ascending node of the corresponding great circle on the celestial sphere.

Modeling the $\omega$ of the hypothetical planet shows that the minimum accumulative $(r-R_d)^2$ is reached at the value of $70^{\circ}$. This means that the angle between the semi-major axes of the two estimated orbits is $25^{\circ}$. The probability that two directions having such an angle are completely different is $(1-\cos 25^{\circ})/2 \approx 0.05$ on the null hypothesis of uniform distribution. By the way, in \cite{guliyev&guliyev2012hypotesis} similar testing was carried out for the nodes and aphelia of periodic comets in order to determine the direction of Jupiter line of apsides. The output value of $\omega$, differed from the true one by $10^{\circ}$. Taking into account inaccuracy of cometary orbits and limited number of them, it can be stated that the agreement between these two values of $\omega$ complies with the hypothesis.

Therefore, if we commit the mean $\omega$, we can assume that the hypothetical planet has the following orbital elements:
$$a = 337\,\mathrm{AU};\:	e = 0.14;\: \omega = 57^{\circ};\: \Omega = 272.7^{\circ};\: i = 86^{\circ}$$

\section{Numerical integrations}

To verify the stability of such an orbit, we integrated it in a heliocentric frame for $10^7$ yr in the past and in the future. For this purpose we used the Bulirsch-Stoer algorithm which is included in the MERCURY package (\cite{chambers1999hybrid}). Within numerical explorations we considered the gravitational forces of the Sun, four Jovian planets and the Planet X, assuming its mass $\sim1M_\oplus$ or $\sim10M_\oplus$.

Furthermore, we selected 53 comets based on their orbital characteristics (inclinations and perihelion latitude) in order to check possible dynamical
path between the comets and the Planet X. The B--S code may handle close encounters with massive bodies when they are involved, thereby we registered any particle that came within 1 Hill radius of the perturber. It is important to note here that a celestial body with masses given above, at such a heliocentric distance ($r>300$ AU), will have a Hill sphere radius $\sim$3--7 AU. The initial orbital elements are obtained from the DE431 numerical theory provided by the JPL ephemeris service\footnote{\href{https://ssd.jpl.nasa.gov/}{https://ssd.jpl.nasa.gov/}}. The simulations were carried out for approximately $10^6$ yr in the past with formal accuracy parameter set $10^{-12}$, was applied. The non-gravitational forces are not considered during the integrations.

Since we do not know the location of the hypothesized planet at the moment, we decided to produce a series of simulations, by changing its mean anomaly on $10^{\circ}$, thereby shifting its position in the orbit almost equidistantly. Thus, we got 36 simulations for given mass.

Our results showed that such an orbit can indeed be sustainable despite the extreme inclination, at least for the period of integration. In addition, we succeeded in detection of the number of close encounters between some comets and the planet. List of these comets will be given below.

\section{Discussion and summaries}

At the beginning of our study, we largely focused on an examination of assumption about concentration of the LPCs perihelion positions around some point or plane (great circle on the celestial sphere). Evidence of one of them rejects the hypothesis of a chaotic perihelia distribution. Our investigation points to the existence of two planes on the celestial sphere, around which concentration of perihelia takes place, consequently, as well as their aphelia. One of these areas is close to the ecliptic region. So it is not the subject of the research. The second one is more crowded with perihelia. Considering the impact of visibility conditions, it might be argued that this area has a significant advantage over the first one. Thus, its existence may be considered as one of the specific features of the LPCs system. The central plane of this area is almost perpendicular to the ecliptic. On the basis of this area instead of the ecliptic, we can see clear patterns in the distribution of certain angular parameters, such as $L',\, B',\, \omega',\, \Omega'_d$. 
It is possible that the existence of such an area can be explained by several factors. In our view, it can be the reason for the existence of a whole family of comets, injected by an unseen massive planet on the distant peripheries of the Solar System. Such a planet could \textquotedblleft generate\textquotedblright two types of comets: some of them have aphelia in the zone of planet motion, others have distant nodes of cometary orbits in the same region. We have focused significant attention on the theoretical search for such groups within our study. It turned out, there is in fact an excess of distant nodes of cometary orbits in the area \eqref{pln1} as in the range 250--400 AU as well as aphelia of a considerable number of comets. We calculated two optimal ellipses considering aphelion positions and distant nodes of the corresponding comets. As a result, the angle between the directions of their semi-major axes is about $27^{\circ}$, which is beyond the scope of randomness. These regularities are well associated with the hypothesis of a massive trans-Neptunian planet. Similar hypotheses have been developed in recent years by other authors, in particular, \cite{batygin2016evidence,trujillo2014sedna,de2014extreme,iorio2014planet}. Looking ahead, we note that our approach to cometary data can be useful for testing such hypotheses.

In our work we have resulted five estimated orbital elements of the prospective planet. But we cannot judge its mass and the exact location in orbit at the moment. Numerical integrations show that the orbit of such a planet should be sufficiently stable, although it is likely that the origin of such an object does not fit into the frameworks of modern cosmogonic theories of the Solar System formation.

Besides that, we explored the orbital evolution of 53 comets up to 1 million years, taking into account the gravitational influence of the unknown planet. In this case, we considered two options for its mass---it could have $1M_{\oplus}$, or exceed it 10 times. For a series of simulations, we vary the mean anomaly of the planet by $10^{\circ}$. It turned out that some comets could have close encounters with the prospective planet shortly before their discovery. Among these comets, we should especially highlight the objects: C/2012 L1, C/2005 N1, C/2006 CK, C/1920 X1, C/2007 D3, C/2010 A4, etc. According to our calculations, a special place is occupied by the comet C/1911 O1, since several centuries before its discovery, it could have a close approach with the planet at a distance of 2--3 AU. Perhaps, it is the comet that could answer the question of where the planet is now. However, it is clear that the planet should have an almost imperceptible daily motion with a very low albedo or absolute magnitude, placing it beyond current detection limits. The proximity of the orbital plane to the galactic plane may potentially complicate the search for such a planet (see also \cite{sheppard2011southern}). 

Anyway, as cometary data increases as well as their accuracy, our calculations, assumptions and conclusions should be verified.

\bibliographystyle{abbrvnat}
\bibliography{winnower_template}

\end{document}